# Fahrzeug Infrastruktur – Kommunikation zur Unterstützung von Fahrerassistenzsystemen

Prof. Dr. Horst Wieker, Arno Hinsberger, Manuel Fünfrocken, Jonas Vogt
Hochschule für Technik und Wirtschaft des Saarlandes

## Abstract

Der Bereich, der Fahrzeug zu Infrastruktur (C2I) Kommunikation hat sich in den letzten Jahren zu einem immer wichtiger werdenden Gebiet im Bereich der C2X Kommunikation entwickelt. Sie ist auf einer Stufe mit der Fahrzeug zu Fahrzeug (C2C) Kommunikation zu sehen und verwendet die gleichen Technologien und Protokolle. Im heutigen Forschungsumfeld ist es Konsens, dass zur Kommunikation der IEEE Standard 802.11p eingesetzt wird. Darauf aufbauend entwickelt das Car 2 Car – Communication Consortium einen einheitliche Protokollstack zur effektiven und effizienten C2X Kommunikation.
Zur Unterstützung von Fahrerassistenzsystemen können sogenannte Roadside Units (RSU) zum Einsatz kommen. Sie sollen durch gezielte Informationsaufarbeitung und Informationsverteilung das Fahrzeug durch zusätzliche Daten in seinen Möglichkeiten erweitern und darüber hinaus Informationen liefern, die allein durch C2C Kommunikation nicht zur Verfügung stehen könnten.

## 1. Einführung

Um Fahrzeuge beziehungsweise Fahrerassistenzsysteme gezielt mit Informationen unterstützen zu können ist es - je nach Anwendungsfall - notwendig, dass die RSU ein detailliertes Wissen über ihre Umgebung hat und gezielt mit Informationen aus den Verkehrszentralen versorgt wird. Zusätzlich dazu muss sie die Informationen der Fahrzeuge auswerten und sowohl an die Verkehrszentralen als auch aggregiert mit den Informationen aus denselbigen wieder zurück an die Fahrzeuge leiten.

Im Zusammenhang mit diesen Informationen muss auf eine Reihe von Aspekten besonders eingegangen werden. Bevor Informationen im Fahrzeug durch Assistenzsysteme weiter verarbeitet werden können und dann auch dem Fahrer angezeigt werden dürfen müssen sie zunächst auf ihre Gültigkeit überprüft werden. Im zweiten Schritt muss das fahrzeugseitige Informationssystem feststellen, ob die entsprechenden Informationen für das Fahrzeug von Relevanz sind.

Die Gültigkeit einer Information ist dabei Abhängig von verschiedenen Kriterien. Da Informationen bei bestimmten Anwendungsfällen nicht überall dort gültig sind, wo sie empfangen werden können, enthalten sie unter Umständen eine Einschränkung auf ein geographisches Gebiet. Dieses Kriterium der geographischen Gültigkeit ist erfüllt, wenn das empfangende Fahrzeug sich in diesem Gebiet befindet.

Des Weiteren gelten Informationen nicht zeitlich unbeschränkt, sondern verlieren zu einem bestimmten Zeitpunkt ihre Gültigkeit. Sollte die Information hiervor nicht aktualisiert worden sein, müssen alle Kommunikationseinheiten diese Information verwerfen und ihre Verbreitung einstellen.

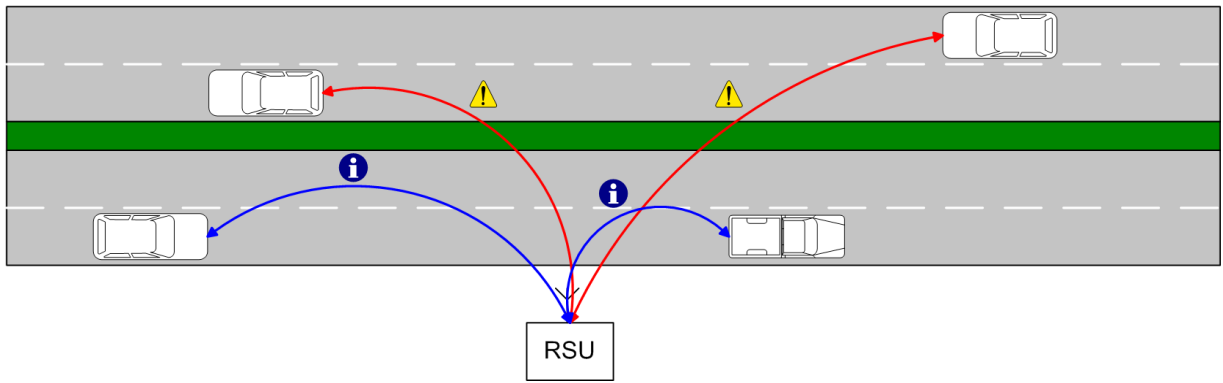

**Abbildung 1:** Unterschiedliche Nachrichten je Richtungsfahrbahn

Eine Nachricht, die gültig ist, muss nicht zwangsläufig für jedes sie empfangende Fahrzeug von Interesse sein. Ob eine Nachricht für den Empfänger von Interesse ist, wird anhand bestimmter Relevanzkriterien entschieden. Dies ist neben der Meldungshäufigkeit, die angibt wie viele Quellen die gleiche Information verteilen, die geographische Relevanz (zone of relevance). Diese besteht entweder aus einer Folge von Koordinaten (der sog. Trace Point Chain) oder aber einer einzelnen Koordinate mit Richtung. Wichtig ist dabei, dass Informationen, die geographisch gültig sind nicht zwangsläufig geographisch relevant sein müssen. Beispiel hierfür sind Fahrzeuge die geographisch benachbart sind, aber in unterschiedlichen Richtungen fahren. Dies ist in Abbildung 1 dargestellt. Ein zusätzliches Kriterium für die Relevanz ist die Art der Information. So sind zum Beispiel Infotainment Nachrichten nur dann für den Empfänger relevant, wenn der entsprechenden Infotainmentdienst aktiv ist.

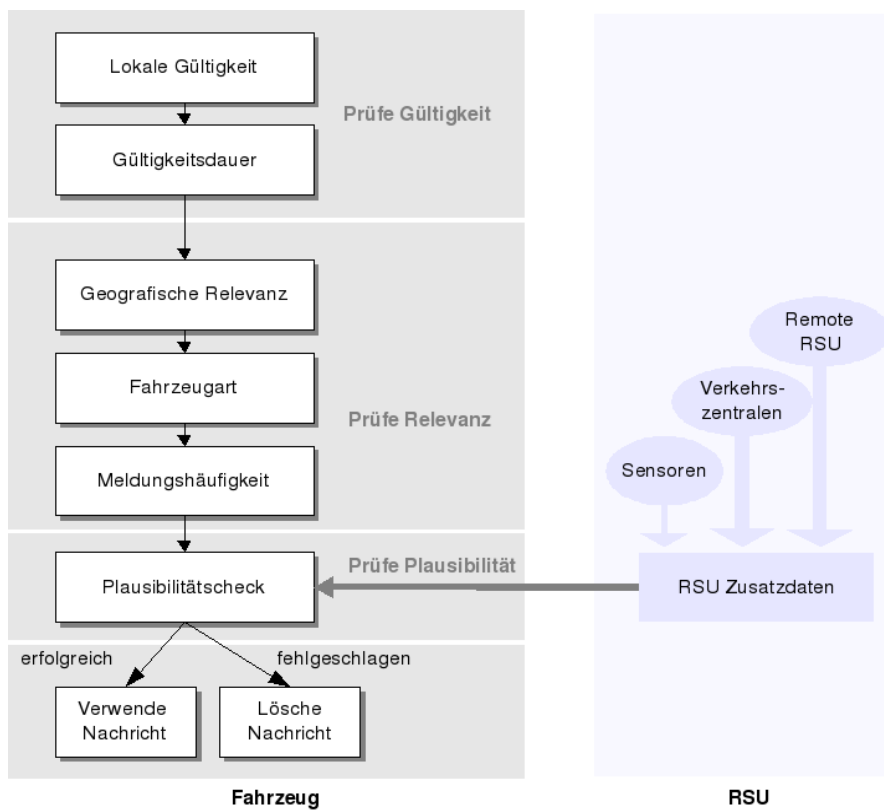

**Abbildung 2:** Plausibiltätsprüfung mit Informationen der Infrastruktur

Alle diese Kriterien treffen jedoch nicht unbedingt eine Aussage über die Plausibilität der Information. Die Überprüfung der selbigen obliegt ausschließlich der Anwendung auf Empfängerseite. So könnte die Plausibilität einer Information über Eisglätte bei einer Umgebungstemperatur von über 20 °C durchaus angezweifelt werden. Dieses Verfahren kann durch zusätzliche Informationen durch die RSU noch verfeinert werden. Diese zusätzlichen Informationen könnte die RSU durch eigene Sensoren, andere RSU oder aus Eingaben von den Verkehrszentralen erhalten.

## 1.1. Nachrichten

Um diese Funktionalitäten zu realisieren hat das C2C CC Basic Services definiert, welche mittels auf IEEE 802.11p basierender Kommunikationstechnologie übertragen werden. Diese Services dienen den auf darüberliegenden Ebenen definierten Anwendungen als Schnittstelle zur Kommunikation. Zurzeit sind zwei dieser Services definiert: CAM (Cooperative Awareness Message) und DEN (Decentralized Environmental Notification Message). Die CAM ist eine periodisch gesendete Nachricht, mit der die Fahrzeuge oder die RSU auf sich aufmerksam machen und dabei noch zusätzliche Informationen wie etwa beim Fahrzeug die Fahrzeugart (PKW, LKW, Bus, Motorrad, …) und die Größe des Fahrzeuges oder die aktuelle Position und Geschwindigkeit bis hin zum Zustand der Scheinwerfer übermitteln. Die RSU weist sich in der CAM mit ihrem Typ aus. Die DEN dient im Gegensatz dazu über warnwürdige Verkehrssituationen (Baustellen, Unfall, liegengebliebene Fahrzeuge, …) zu informieren.

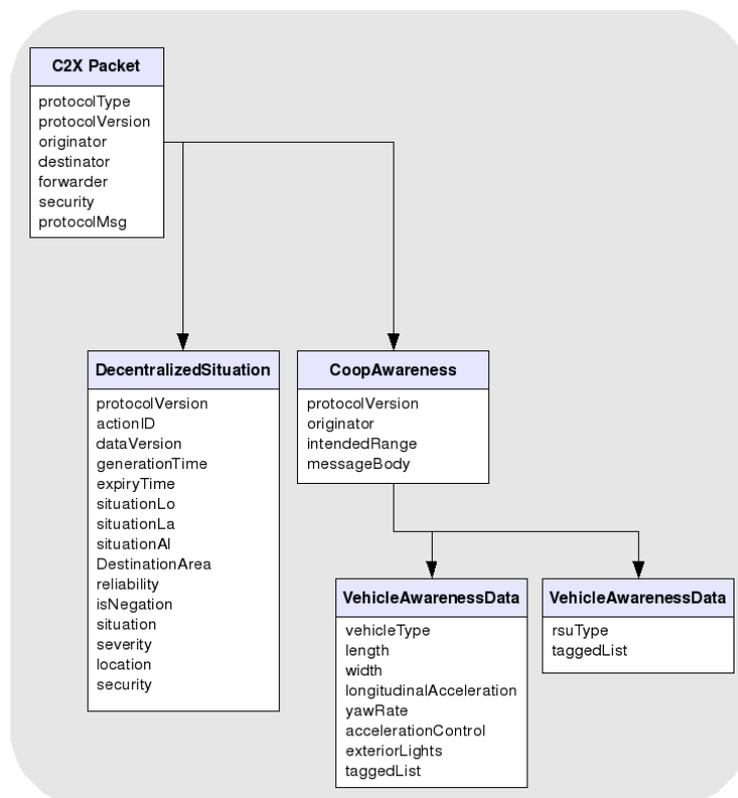

**Abbildung 3:** CAM und DEN im Detail (C2C-CC Demo 2008)

Mit diesen definierten Nachrichten sind bereits heute schon eine Vielzahl verschiedener Szenarien realisierbar. Dabei reicht das Spektrum von Warnungen vor möglichen Kollisionen an unübersichtlichen Kreuzungen, über Informationen zu Baustellen bis hin zu Warnungen vor sich nähernden Rettungsfahrzeugen.

Abbildung 4 verdeutlicht das Zusammenspiel von Anwendungen, Protokollelementen und dem Übertragungssystem. Wie bereits angedeutet, nutzen Anwendungen (Applications) die Basisdienste (Basic Services) zum Kommunizieren. Die Basic Services werden sich aus einer Reihe von Diensten zusammensetzen die wiederum das Kommunikationssystem (Communication System) nutzen, um Informationen zu übertragen.

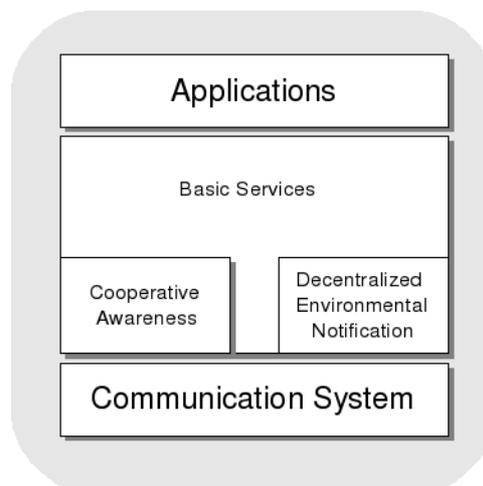

**Abbildung 4:** Aufbau des C2C-CC Demo 2008 Protokoll Stacks.

## 2. RSU

Die RSU ist der zentrale Bestandteil der C2I Kommunikation. Sie stellt das Verbindungsglied zwischen Fahrzeugen und Infrastruktureinrichtungen (z.B. Verkehrszentralen) dar.

Die RSU besteht aus einem 802.11p Kommunikationsmodul und einer Anwendungsplattform, die die RSU-seitigen Funktionsanteile beinhaltet. Diese RSU Software hat sehr unterschiedliche Aufgaben. Sie ist zum einen für die Kommunikation zur Zentrale zuständig. Diese Kommunikation ist dabei zumeist kabelgebunden und verbindungsorientiert. Auf der andern Kommunikationsseite existiert eine verbindungslose 802.11p Ad-Hoc Kommunikation. Ein Problem stellt hier dar, dass die proprietären Protokolle die für die Infrastrukturkommunikation benutzt werden nur höchst selten Bandbreitenbeschränkungen berücksichtigen. Dies führt dazu, dass die zu übertragenen Daten zum Teil sehr ineffizient Kodiert sind. Die RSU muss daher in der Lage sein zwischen diesen beiden Welten mit unterschiedlichsten Protokollen und Anforderungen an die Kommunikation zu übersetzten.

Zusätzlich hat sie die Aufgabe eine intelligente Nachrichtenverteilung und -aggregation durchzuführen, da zum einen eine einfache Weiterleitung der Nachrichten nicht den in diesem Umfeld an die RSU gestellten Anforderungen an eine effiziente Bandbreitennutzung gerecht werden kann, zum anderen lassen sich Nachrichten aus dem Infrastrukturbereich und der Ad-Hoc Kommunikation nur selten direkt aufeinander abbilden.

Ihre intelligente Funktionalität erlaubt es der RSU, ihr Kommunikationsverhalten den gegebenen Verkehrsverhältnissen anzupassen und dynamisch auf deren Änderungen zu reagieren. Je nach Anforderungen können verschieden Funktionalitäten auf der RSU installiert werden, wodurch den unterschiedlichen Rahmenbedingungen von Autobahn oder innerstädtischen RSU Rechnung getragen werden kann.

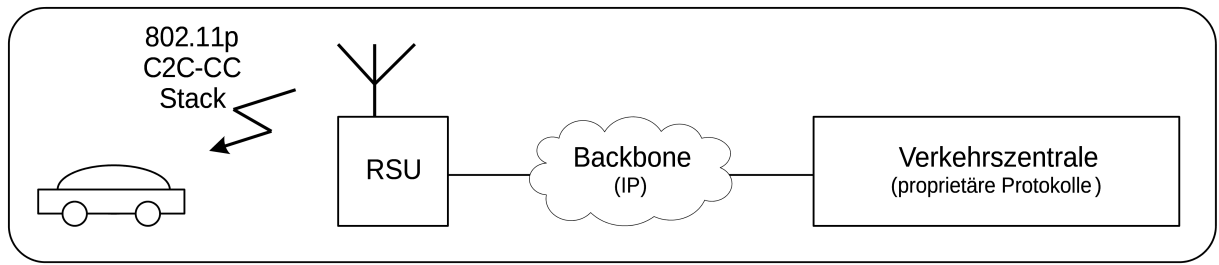

**Abbildung 5:** Die RSU in der C2I Kommunikation

Abbildung 5 zeigt schematisch die gesamte, bidirektionale Kommunikationskette vom Fahrzeug zur Verkehrszentrale.

Die RSU ist im Gesamtsystem als die zentrale Schnittstelle des Traffic Management zu betrachten. Sie erhält aus verschiedensten Quellen - neben den Fahrzeugen - Informationen über aktuelle Baustellen, Wetterbedingungen, geplante Großveranstaltungen, geschaltete Schilderbrücken, Lichtsignalanlagen, etc. Alle diese Daten werden in der RSU aufbereitet und zu sinnvollen und möglichst effizienten Nachrichten aggregiert.

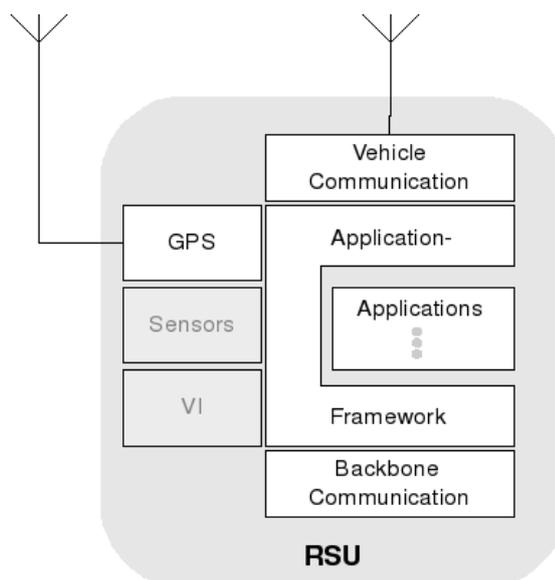

**Abbildung 6:** Aufbau der RSU (schematisch)

## *2.1. Übergeordnete Aufgaben der RSU*

Die Aufgaben einer RSU lassen sich in folgende Bereiche gliedern:

- **Verkehrssicherheit** Ist der wichtigste Punkt in diesem Zusammenhang. Die RSU kann beispielsweise - aus Informationen von den Verkehrszentralen - ein detailliertes

Bild der aktuellen Verkehrssituation. Mit Hilfe dieser Informationen kann die RSU dem Fahrzeug bzw. genauer dem Fahrerassistenzsystem Angaben über die momentane Straßenlage bereit stellen, auf die dieses ohne das Vorhandensein einer RSU nicht zurück greifen könnte. Auf diese Weise kann z.B. die Funktionalität von Querregelungsassistenten deutlich erweitert und verbessert werden.

Die RSU verfügt über einen internen Speicher in dem sie empfangenen Nachrichten während ihrer Gültigkeit vorhält, um sie bei bedarf an Fahrzeuge weiterzuleiten. Durch diese Store and Forward genannte Technologie kann die RSU Nachrichten (auch solche, die eigentlich für die direkte Fahrzeug-zu-Fahrzeug Kommunikation gedacht sind) an Fahrzeuge weiterleiten, wenn die aktuelle Verkehrsdichte aufgrund der geringen Zahl von Fahrzeugen eine direkte C2C Kommunikation nicht oder nur sehr sporadisch ermöglicht. Dies ist vergleichbar mit der Situation, die sich bei der Markteinführung der C2X Kommunikation ergibt, da zu diesem Zeitpunkt nur eine geringe Anzahl von Fahrzeugen mit Kommunikationseinheiten ausgestattet sein wird.

- **Verkehrseffizienz im Microbereich** Einem Fahrerassistenzsystem kann durch Informationen aus den Verkehrszentralen ermöglicht werden, dem Fahrer gezielt Angaben über die aktuelle Verkehrslage, Reisezeiten, Stauwarnungen und Umleitungsempfehlungen anzuzeigen.

- **Infotainment** Neben der Unterstützung von Fahrerassistenzsystemen bietet die C2X Kommunikation auch die Möglichkeit, dem Fahrer Infotainmentdienste zur Verfügung zu stellen. Diese zusätzlichen Funktionalitäten werden getrennt von den Sicherheitsinformationen übertragen um diese nicht zu stören. Beispielsweise könnten RSU an den Zufahrtsstraßen so programmiert werden, dass sie ankommenden Fahrzeugen Informationen über Veranstaltungen, Parkplätze, Sehenswürdigkeiten, Gastronomieempfehlungen und generelle Informationen zur Verfügung stellen. Das Assistenzsystem im Fahrzeug kann vom Fahrer dann so konfiguriert werden, dass es lediglich die gewünschten Informationen anzeigt.

## *2.2. Standalone RSU*

Nicht immer benötigt eine Roadside Unit eine Anbindung an die Verkehrszentrale um Fahrzeuge mit hilfreichen Zusatzinformationen versorgen zu können. Gerade der Unfallschwerpunkt Autobahnbaustelle bietet Anwendungsmöglichkeiten für autarke RSU. Diese sog. Standalone RSU kann in einem Baustellenanhänger untergebracht werden. Diese RSU kann Informationen über Durchfahrzeiten, Baustellengeometrie (Spurverlauf) und ähnliches an einfahrende Fahrzeuge übermitteln. Wird zusätzlich zu der ersten RSU, der sog. Zulauf RSU, eine zweite RSU am Baustellenausgang (Ablauf RSU) eingesetzt, können diese die Informationen über Durchfahrzeiten und Baustellengeometrie selbstständig ermitteln. Hierzu werden die Fahrzeuge durch die Zulauf RSU aufgefordert, die Durchfahrzeit und den Verlauf der zurückgelegten Strecke zu messen. Diese Informationen geben sie bei der Ablauf RSU ab. Diese übermittelt die aggregierten Informationen an die Zulauf RSU, welche daraufhin ihre Angaben über Baustellengeometrie und Durchfahrzeiten aktualisieren und verfeinern kann. Hierdurch werden den ankommenden Fahrzeugen ständig aktuelle und sehr genaue Daten zur Verfügung gestellt, ohne dass Baustellenpersonal in der Programmierung der RSU eingewiesen werden muss.

Sind genaue Daten über Baustellengeometrie und Streckenverlauf durch Vermessungen vorhanden, können diese ankommenden Fahrzeugen zur Verfügung gestellt werden. Dies erlaubt z.B. den Querregelungsassistenten von LKW den Fahrer während der Fahrt durch die Baustelle zu unterstützen und somit ein Schlingern zu vermindern.

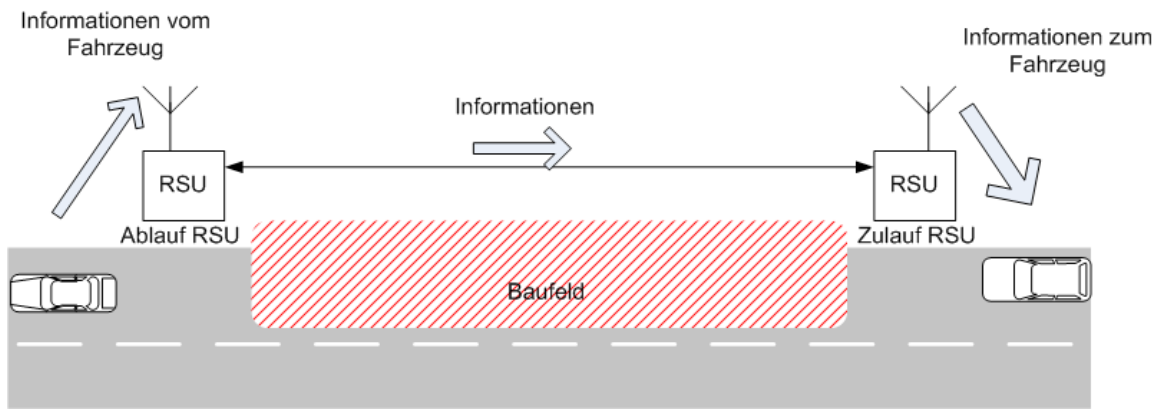

**Abbildung 7:** Die Standalone RSU an der Baustelle

## 3. Anwendungen der C2I Kommunikation

Die besondere Stärke der C2I Kommunikation in Kombination mit der C2C Kommunikation liegt zum einen in der Erhöhung der Plausibilität von Nachrichten zum anderen in der Hilfestellung bei übertragungstechnisch schwierigen Situationen.

**Erhöhung der Plausibilität**

Ähnlich wie bei der Relevanz kann die Plausibilität einer Information durch die Häufigkeit einer Meldung steigen. Insbesondere kann dies durch zusätzliche Informationen von einer RSU erfolgen. Dementsprechend könnte z.B. die Plausibilität einer Glatteismeldung auf einer Brücke dadurch erhöht werden, dass eine RSU in der Nähe für dieses Gebiet eine generelle Glätte Information sendet, ähnlich wie es heute durch das Verkehrzeichen 1007-30 „Gefahr unerwarteter Glatteisbildung" geschieht. Analog dazu kann dies natürlich auch bei anderen Gefahrenwarnungen durchgeführt werden, wie z.B. Staugefahr.

**Unterstützung der Kommunikation**

Eine weitere Schlüsselrolle kommt der C2I Kommunikation bei der Weiterleitung von Information an besonders sensiblen bzw. gefährlichen Stellen bzw. bei ausbreitungstechnisch ungünstigen Umgebungen zu. Eine Situation bei der alle diese Kriterien zusammen kommen ist die urbane Kreuzung: So könnte eine RSU im Kreuzungsbereich eine Staumeldung, die wegen Abschattung nur in einer Straße empfangen werden kann, in die anderen Straßen weiterleiten, so dass Fahrerassistenzsysteme noch vor Einfahrt in die Kreuzung Umleitungsempfehlungen geben können.

**Unterstützung von Fahrerassistenzsystemen**

Des Weiteren bietet die RSU eine Möglichkeit, die Arbeitsweise von Kollisionswarnsystemen zu verbessern. Indem sie CAM Nachrichten aller Fahrzeuge, die aus einer Richtung kommen sammelt und Informationen hierüber an die Fahrzeuge, die aus anderen Richtungen kommen und sich daher im Funkschatten befinden, abgibt. Hierdurch haben alle Fahrzeuge bessere Kenntnis über die momentane Verkehrssituation, ohne dass es zu einer Überlastung auf dem Kommunikationskanal durch die vielen CAM kommt.

# 4. Zukunft C2X

## 4.1. Erweiterung der Basic Services

Die Planungen zur Erweiterung der Funktionalitäten gehen zurzeit in die Richtung der Erweiterung der Basic Services um die Bereiche Geographical Services. In diesem Themenbereich soll die sog. Local Dynamic Map weitere Möglichkeiten bei der Traffic Efficiency aufzeigen. Bei diesem Thema sind allerdings noch viele Fragen offen. Angefangen bei der Frage welche Informationen werden für welche Situation z.B. an einer urbanen Kreuzung von Interesse sein? Wie detailliert müssen die Daten über die spurabhänige Fahrzeugdetektion sein, damit dadurch eine Effizienzsteigerung bei der Verkehrssteuerung möglich ist? Bis zu der Frage welche Informationen, wie etwa die Reisezeit, gewonnen werden können, wie eine Ampelsteuerung optimiert werden kann und in wieweit Warnungen vor Unfällen mit Fußgängern, besonders Kindern, vermieden werden können.

Sind diese Fragen beantwortet, kann damit begonnen werden zu klären, inwieweit Funktionsanteile von kooperativen Anwendungen auf die RSU ausgelagert werden können. Da die RSU zum einen über eine höhere (und leichter erweiterbare) Rechenkapazität als die Fahrzeuge verfügt und zum anderen die berechneten Ergebnisse mehreren Fahrzeugen zur Verfügung stellen kann, ließen sich die vorhandenen Informationen und Übertragungskapazitäten deutlich effizienter nutzen.

## 4.2. Sensorische Umfelddaten

Weiterhin können RSU mit state-of-the-art Sensoren ausgestattet werden, die es ihnen erlauben neben Fahrzeugen ohne Funkaustattung auch Fußgänger oder Radfahrer zu detektieren. Diese Informationen bilden zusammen mit allen anderen gesammelten Daten und Informationen aus den Verkehrszentralen eine umfassende Darstellung der Kreuzungssituation mit deren Hilfe die RSU mögliche Gefahrensituationen frühzeitig erkennen kann. Hierbei muss die RSU nicht unbedingt eine Gefahrenwarnung an die Fahrzeuge senden, sondern sie kann durch die gezielte Informierung von Fahrzeugen dafür sorge tragen, dass diese die Gefahr selbsttätig erkennen können. Die RSU stellt somit fest, welche zusätzlichen Informationen wann für ein Fahrzeug von Interesse sein könnten. Die Assistenzsysteme in den Fahrzeugen werten diese selbsttätig aus und entscheiden wie reagiert werden muss.

Diese vielen verschiedenen Dienste müssen natürlich auch verwaltet werden. Zum einen innerhalb der RSU selbst, zum anderen wird nicht jedes Fahrzeug jeden Dienst unterstützen oder benötigen. Daher müssen in Zukunft Systeme entwickelt werden, die es Fahrzeugen erlauben, nur bestimmte Dienste anzufordern. Dienste, die zurzeit nicht benötigt werden, könnten dann von der RSU pausiert werden um die ohnehin stark beschränkte Bandbreite auf der Luftschnittstelle nicht unnötig zu belasten.

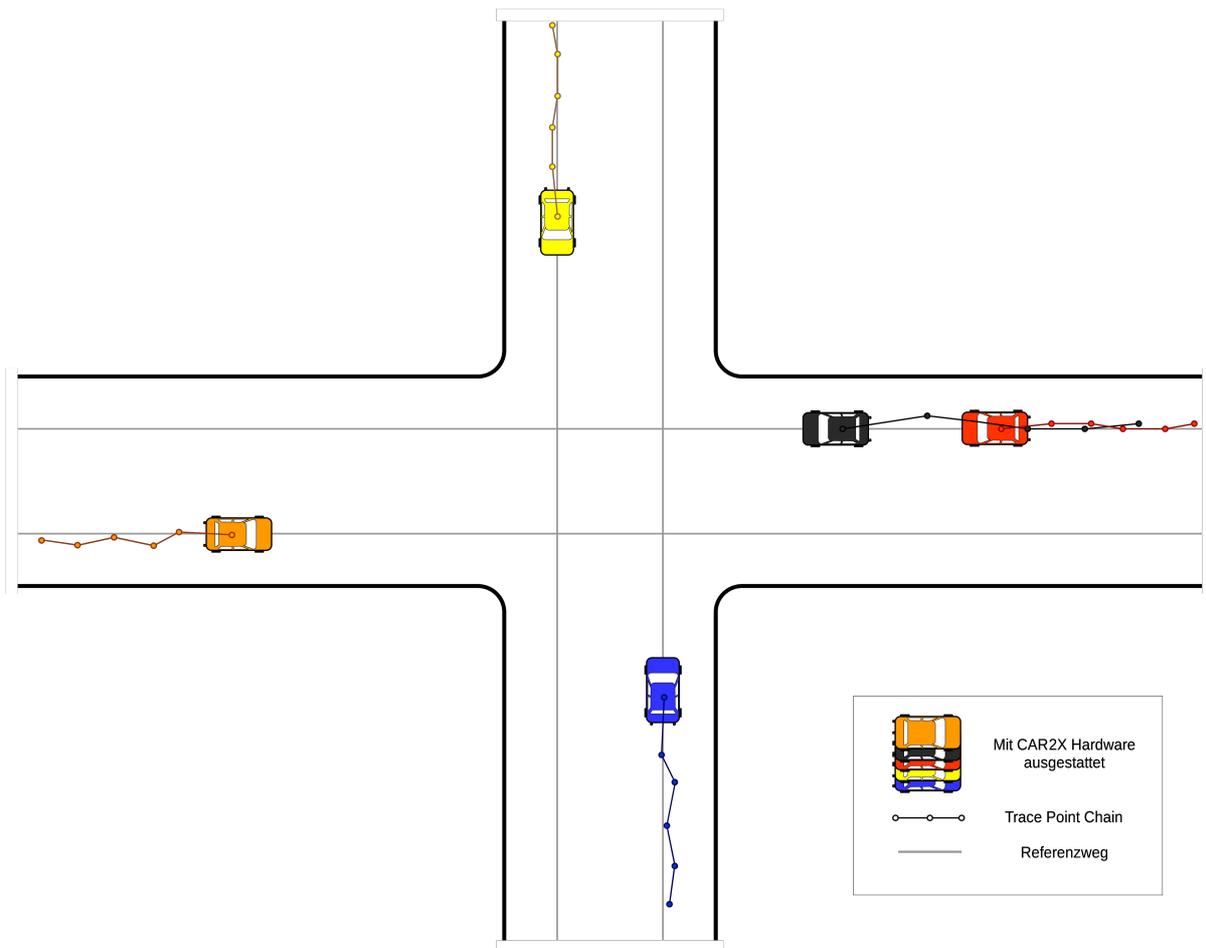

**Abbildung 8:** Die Kreuzung mit dem einfachen Datenmodell und den C2C Informationen

Die in Abbildung 8 dargestellte einfache Verkehrssituation an einer Kreuzung entspricht in etwa der Darstellung im Fahrzeug. Das Fahrzeug kennt die Position anderer Fahrzeuge, die grobe Geometrie der Kreuzung und seine eigene Position.
Das Modell, das die RSU verwenden kann, ist hier schon deutlich besser. Wie in Abbildung 9 zu sehen besitzt sie detaillierte Informationen über den Aufbau der Kreuzung wie unter Anderem die Anzahl der Spuren, die Position von Ampeln und Verkehrszeichen sowie die Referenzwege. Außerdem könnte sie auch Informationen über Verkehrsteilnehmer haben, die nicht an der C2I Kommunikation beteiligt sind, wie z.B. Fahrräder, Fußgänger oder unausgestattete Fahrzeuge. Dies alles kann zu einem komplexen Modell der Kreuzung zusammengesetzt werden. Bei Bedarf können dann Teile dieses Modells an Fahrzeuge übermittelt werden um deren Einschätzung der Verkehrssituation zu verbessern.

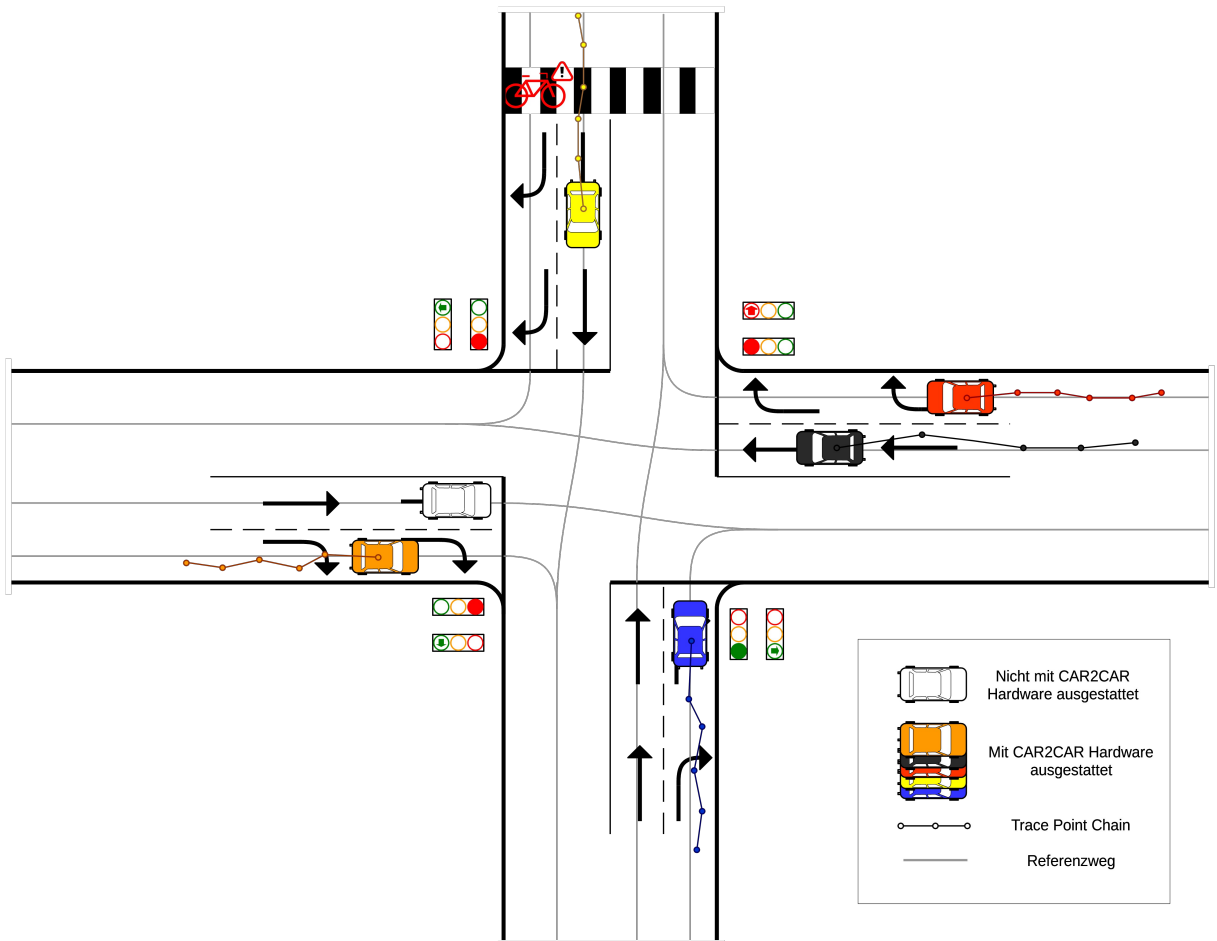

**Abbildung 9:** Die Kreuzung mit komplexem Datenmodell und Informationen der C2I Kommunikation.

## 5. Schlusswort

Wie bereits beschrieben, können Fahrerassistenzsysteme schon durch die Basic Services CAM und DEN profitieren. Größere Vorteile werden aber die effiziente Nutzung der deutlich beschränkten Bandbreite der Luftschnittstelle und der besseren Informierung der Fahrzeuge bieten, wie sie die RSU ermöglichen kann. Zu klären bleibt, welche Informationen über die Kreuzung notwendig sind um eine verkehrliche Wirkung zu erzielen und in welcher Genauigkeit diese Informationen vorliegen müssen. Außerdem muss geklärt werden, wie diese Informationen strukturiert und kodiert werden können um sie möglichst effizient zu übertragen.

# Literaturverzeichnis